\documentclass[12pt,preprint]{aastex}
\newcommand{\kms}{km\,s$^{-1}$}
\newcommand{\cm}{$\mathrm{cm^{-1}}$}
\shorttitle{Discovery CH in $\eta$\ Car} \shortauthors{Verner et
al.}

\received{2004 September 16}
\begin{document}

\title{Discovery of CH and OH in the $-$513~km~s$^{-1}$ Ejecta of
Eta Carinae}

\author{ E.~Verner\altaffilmark{1,2,3},
F.~Bruhweiler\altaffilmark{1,2},
K.~E.~Nielsen\altaffilmark{1,2}, T.~R.~Gull\altaffilmark{1},
G.~Vieira Kober\altaffilmark{1,4}, and
M.~Corcoran\altaffilmark{5}}

\altaffiltext{1}{Exploration of the Universe Division, Code 667,
Goddard Space Flight Center, Greenbelt, MD 20771} \altaffiltext{2}
{IACS/Dept. of Physics, Catholic University of America,
Washington, DC 20064} \altaffiltext{3} {Dept. of Engineering,
Architecture \& Aerospace Technology, University of the District
of Columbia, Washington, DC 20008} \altaffiltext{4}{Science
Systems and Applications, Inc, Lanham, MD 20706}
\altaffiltext{5}{Universities Space Research Association, 7501
Forbes Blvd, Suite 206, Seabrook, MD 20706 and the Exploring the
Universe Division, NASA Goddard Space Flight Center, Greenbelt,
MD 20771}

\email{kverner@fe2.gsfc.nasa.gov, fredb@iacs.gsfc.nasa.gov,
nielsen@stis.gsfc.nasa.gov, theodore.r.gull@nasa.gov,
gvieira@stis.gsfc.nasa.gov, corcoran@barnegat.gsfc.nasa.gov}

\begin{abstract}

The very massive star, Eta Carinae ($\eta$ Car) , is enshrouded in
an unusual complex of stellar ejecta, which is highly depleted in
C and O, and enriched in He and N. This circumstellar gas gives
rise to distinct absorption components corresponding to at least
20 different velocities along the line-of-sight. The velocity
component at $-$513~\kms{} exhibits very low ionization with
predominantly neutral species of iron-peak elements. Our
statistical equilibrium/photoionization modeling indicates that
the low temperature (T = 760 K) and high density
(n$_\mathrm{H}$$\sim$10$^{7}$ cm$^{-3}$) of the $-$513~\kms\
component is conducive to molecule formation including those with
the elements C and O. Examination of echelle spectra obtained with
the Space Telescope Imaging Spectrograph (STIS) aboard the {\it
Hubble Space Telescope (HST)} confirms the model's predictions.
The molecules, H$_2$, CH, and most likely OH, have been identified
in the $-$513 \kms{} absorption spectrum. This paper presents the
analysis of the {\it HST}/STIS spectra with the deduced column
densities for CH, OH and C I, and upper limit for CO. It is quite
extraordinary to see molecular species in a cool environment at
such a high velocity. The sharp molecular and ionic absorptions in
this extensively CNO- processed material offers us a unique
environment for studying the chemistry, dust formation processes,
and nucleosynthesis in the ejected layers of a highly evolved
massive star.

\end{abstract} 

\section{Introduction} Eta Carinae is a Luminous Blue
Variable (LBV) of extremely high mass (M$_{*} \sim$ 100$_\sun$) in
a
late phase of its evolution. Most of the nebulosity enshrouding
$\eta$ Car is a product of earlier ejection  events. The bipolar
structure, known as the Homunculus, was created during the great
eruption in the 1840's, while a similar structure, the Little
Homunculus \citep{Ish03}, resulted from a less significant
eruption in the 1890's. Eta Car has undergone extensive mass loss
and has a current mass loss rate of  \.M$\sim$10$^{-3}$
M$_\sun$yr$^{-1}$ (Hillier et al. 2001).

\indent The stellar photosphere, as well as nebulosities both
close to $\eta$ Car and beyond the Homunculus indicate significant
CNO-processing (Hillier et al. 2001). Emission line spectra of the
Weigelt Blobs, B and D, stellar ejecta at distances of $\sim$
200-600 AU indicate
 extensive CNO-processing (Verner, Bruhweiler \& Gull 2005).
Even at much larger distances, optical/UV
emission spectra show N-enhancement within the soft X-ray emitting
shell surrounding $\eta$ Car (Davidson et al. 1986; Smith
\& Morse 2004).

\indent Observations using {\it HST}/STIS of the central emission
core containing the stellar source of $\eta$ Car  reveal a complex
of narrow circumstellar absorption lines representing at least 20
distinct velocity components in the line-of-sight toward $\eta$
Car. The high velocity component at $-$513 \kms{} arises  in a
region of low temperature, T=760 K (Gull et al. 2005). This
resolved component is typified by both neutral and singly-ionized
elements.  A curve-of-growth analysis provided level populations
and an excitation  temperature which were incorporated into
statistical equilibrium/photoionization modeling similar to that
used in \citet{Ver02}.
Modeling of this cool component yielded a characteristic distance
to the central star of
$\log$\,(R/cm)$\sim$17.4 for a density of
$\log$\,(n$_\mathrm{H}$/cm$^{-3}$) $\sim$ 7. Another result of the
modeling  was the prediction of
observable molecules in this fast moving filament.

 \indent This paper presents
the observationally derived column densities for CH and OH and
upper limits for other relevant molecular and atomic species in
the $-$513 \kms{} component. Examination of high resolution {\it
HST}/STIS MAMA spectra acquired with the E230H grating reveals
definite detections of six absorption lines from CH and two
probable OH line detections in the wavelength region between 3070
and 3150 \AA. In addition, data at shorter wavelengths yield
column densities for atomic C~I, and a column density upper limit
for CO. Then, comparisons between modeling and observed
column densities set further constraints on the physical
conditions for the $-$513 \kms{} component. Over 800 H$_2$ Lyman
absorption lines have been identified in the $-$513 \kms{} velocity  component,
and discussed elsewhere \citep{Nie05}.

 \indent The observed Ti II, C I and CH column densities
serve as a basis for photoionization modeling that leads to
predictions for other molecules in the high velocity ejecta. These
results require carbon and oxygen deficiencies, in agreement with
CNO-processing, to explain the observed molecules. Due to large
uncertainties in defining  parameters in the models, we can only
estimate the possible range of physical conditions that are
responsible for molecule formation.

\section{ Observations}

\indent The data for this paper were obtained as a part of several
$\eta$ Car {\it HST} Programs using the high spatial and spectral
resolution capabilities of STIS.  The STIS observations were
obtained with the E140M  (1150$-$1730 \AA) and the E230H
(2885$-$3160 \AA) gratings providing a spectral resolving power of
R=45\,800 and 114\,000, respectively. Additional information about
these observations is presented in Table~\ref{obs}.

\section{Molecular and Neutral Atomic Species
in the $-$513 \kms\ Component}

\subsection{CH and OH}
\indent The molecule CH is definitely present in the circumstellar
environment of $\eta$ Car. Figure~1 shows a portion of the E230H
spectrum with identified CH lines at $-$513 \kms{}. Table~\ref{CH}
presents both measured and upper limits for equivalent widths for
CH transitions of the  C$^{2}$$\Sigma$$^+$$-$X$^2$$\Pi$ system
identified in the spectrum. Table~2 also includes observed and
laboratory wavelengths for each CH transition, as well as measured
velocity, lower energy level of the transition, and gf-values from
\citet{KurCH}. We have searched for CH absorption lines from
levels with excitation energy up to 579 cm$^{-1}$ above ground,
but no detectable absorption is found from any levels with
energies above 18 cm$^{-1}$ with a $\sim$ 2 m\AA{} detection limit
for the individual lines. The reader is referred to Figure~2 of
\citet{Lien84} for the term structure of CH including
$\Lambda$-doubling. The absence of absorption from energy levels
higher than 18 cm$^{-1}$ implies a very low excitation temperature
for CH, T $\leq$ 30~K. This is significantly below the 760~K
inferred from the level populations for Ti II \citep{Gul05}.

\indent For all derived column densities, we assume the features
lie on the linear curve-of-growth. Comparisons with the
curve-of-growth for a b-value of 2.1~km~s$^{-1}$, derived for the
$-$513 km s$^{-1}$ component from the Ti~II lines, indicate that
this assumption is reasonable. The average velocity of the
observed CH lines is $-$513.2\,$\pm$\,1.7 \kms{} which is in good
agreement with the corresponding value derived from the Ti II
lines. Based on the results (Table~2), the total CH column
density is N$_\mathrm{CH}$=5.4\,$\times$\,10$^{13}$ cm$^{-2}$. \\
\indent Although such studies are beyond the scope of this paper,
we further note that the CH lines in the $-$513 km s$^{-1}$
component show  temporal variability between January 2002
and September 2003. We have used only the January 2002 data to
derive column densities.

 \indent Two weak OH lines,
$\lambda\lambda$3079, 3082, are identified in the $-$513 km
s$^{-1}$ absorption spectrum (Figures 2 and 3) and presented in
Table~\ref{OH}. To confirm the presence of these spectral
features, a spectrum obtained in September 2003 has been used for
comparison. Line identification of the OH features was
accomplished using molecular data presented in Roueff (1996). No
weaker OH transitions are observed. Based upon the equivalent
widths of the OH $\lambda\lambda$3079, 3082 features, 1.9 $\pm$
1.0 and 2.3 $\pm$ 1.0 m\AA, respectively, we adopt an OH column
density of N$_\mathrm{OH}$ = (3 $\pm$ 1) x 10$^{13}$~cm$^{-2}$.
Given the weakness of these two lines, even though there is a good
velocity correlation, we consider OH a probable detection in the
circumstellar gas in the line-of-sight to $\eta$ Car.

\subsection{Atomic Carbon, C I}
\indent We have identified C I lines rising from all three
fine-structure levels of the ground term, in the $-$513 \kms{}
component (see Figure 4). Neither the presence of C I, nor its
excitation should be surprising. Yet, the excitation of the
fine-structure levels can lead to additional constraints on the
physical conditions in this component. Because most of these C I
transitions are blended and approaching saturation, we have
derived approximate C I column densities based upon weaker
features, which have resolved C I contributions. The results for
specific lines from the differing fine-structure levels in the
ground configuration are given in Table~\ref{CI}. The C I, C
I$^*$, and C I$^{**}$ refer to transitions originating from the
zero, 16.4, and 43.4 cm$^{-1}$ levels, respectively. The E140M
spectrum, in which the C~I lines are present, has lower spectral
resolution (R=45\,800) than the E230H. We have used the best
resolved components of the transitions of C I to estimate column
densities. Several lines appear to be unblended with no
discernable overlap with other transitions, namely C~I
$\lambda\lambda$1328.833, 1657.379, 1658.121. By adopting a
curve-of-growth corresponding to a b = 2.1~km~s$^{-1}$ as found
for the Ti II (Gull et al. 2005), we derive log(N$_\mathrm{C\,
I}$/cm$^{-2}$)=13.8, log(N$_\mathrm{C\,I^{*}}$/cm$^{-2}$)=13.6,
and log(N$_\mathrm{C\,I^{**}}$/cm$^{-2}$)=14.0. Here, we
specifically bias the result based upon the weaker C I lines since
the deduced column densities are less sensitive to saturation
effects. The average velocities of these lines are
514.1~$\pm$~1.7~km~s$^{-1}$. Summing these column densities gives
a total C I column
density of N$_\mathrm{C\,I(tot)}$=2$\times$10$^{14}$ cm$^{-2}$.\\
\indent We have further used the ratios f$_1$=N$_\mathrm{C\,I^*}$
/N$_\mathrm{C\,I(tot)}$ and
f$_2$=N$_\mathrm{C\,I^{**}}$/N$_\mathrm{C\,I(tot)}$ to estimate
the pressure, P/k = nT (cm$^{-3}$K) for the neutral carbon
absorbing region \citep[e.g.][]{Jen02,JS79}. The resulting
f$_1$=0.3 and f$_2$=0.5 are close to their LTE values. With the
uncertainties, we find that log(nT/(cm$^{-3}$K))~$\geq$~6, given
the temperature, 760K, based upon the analysis of the Ti II lines
\citep{Gul05}.

 \indent With the exception of  H$_2$, the most
abundant molecular species in the interstellar medium is typically
CO. However, C and O may be depleted up to 100 times in the
$\eta$~Car environment \citep{Duf97,Hil01,VBG05}. Negligible CO
may be present in the ejecta. Nevertheless, we did search for CO.
The CO features from the A$^1$$\Pi$ $-$ X$^1$$\Sigma$$^{+}$ bands
(0$-$0 through 7$-$0), fall in the spectral range 1300-1520 \AA.
Based upon f-values the strongest band is the (0$-$0) band near
1510 \AA\ and the bands get weaker toward shorter wavelengths.
Using the CO data from \citet{MN94}, we searched for  the
strongest lines in these bands at velocities within $\pm$ 5 \kms\
of $-$513 \kms\ for each transition wavelength. In all cases, we
have equivalent width upper limits of W$_{\lambda}$ $\leq$ 3m\AA{}
for the R(0) lines of the (0$-$0) and (1$-$0) bands. Likewise,
features from other, weaker bands yielded non-detections.  Using
the f-values for the R(0) transition in the (0$-$0) band to impose
an upper limit, we find that $\log$
(N$_\mathrm{CO}$/cm$^{-2}$)$\leq$ 12.95. This upper limit could be
slightly higher depending upon how the rotational levels are
populated. The CO upper limit is 5-6 times less than the derived
column of CH and 3 times less than our possible detection of OH.
This is in sharp contrast with what is found in the interstellar
medium, where in the case of the line-of-sight to $\zeta$ Oph and
other stars, the CO is roughly 100 times more abundant than CH
\citep[c.f.][]{vDB86}.

\subsection{Molecular Hydrogen, H$_{2}$}
\indent Molecular hydrogen has been observed in the $-$513 \kms{}
ejecta (Smith 2002; Neilsen, Gull \& Viera Kober 2005). As stated
earlier, over 800 features arising from the Lyman bands have been
identified. The absorption lines from the Lyman bands, as seen by
{\it HST}/STIS, are from high vibrational and rotational states.
Ground state transitions are inaccessible in the {\it HST}
wavelength range. We have made a rough estimate of the H$_2$
column density based on the Lyman absorption in the $-$513
km\,s$^{-1}$ component. For T = 760 K, we estimate a lower limit
to the H$_2$ column density to be $\sim$ 10$^{16}$ cm$^{-2}$ based
on H$_2$ Lyman absorption from higher energy states (E$>$11\,000
\cm ) and assuming Boltzmann statistics. However, the relative
populations maybe altered significantly from LTE conditions for
several reasons: 1) by radiative rates (Sternberg \& Dalgarno
1981); 2) contribution from additional hidden cold (T$<$760K)
component. The excitation conditions for H$_{2}$ are beyond the
scope of this paper and will be addressed elsewhere. A total
column density for H$_{2}$ in the range of $\sim$ 10$^{20}$
cm$^{-2}$ would not be unreasonable. In any event, any estimate of
 H$_{2}$ total column density is highly uncertain.

\section{Physical Conditions in the Ejecta}
\indent Due to the complex circumstellar environment of
$\eta$\,Car, it is difficult to develop a unique photoionization
model describing the whole system. The most effective way is to
model separately each velocity component to obtain the physical
conditions in the gas. The binary system may contain a
very luminous B star and a less luminous O or WN star, where the
radiation field in the surrounding nebula can vary dramatically
over orbital phase \citep{VBG05}. In many distinct ejecta the
excitation conditions change over the  5.54 year period
\citep{Dam96,Cor01}. Because of the constant strength and low
ionization of the atomic absorption in the $-$513 km s$^{-1}$
component, we assume that the most important excitation and
ionization source for ejecta at large distance is the radiation
field of the primary B-star. \cite{Ver02} found that a radiation
field of T$_\mathrm{eff}$=15\,000 K  for the central star for the
distance of the Weigelt B and D Blobs, reproduced the presence and
strengths of strong Fe~II and [Fe II] emission lines in these
nebular condensations.

\indent We have attempted to explain the origin of CH, CO and OH
molecules of high velocity ejecta quantitatively. We have used the
photoionization code Cloudy version 96 (Ferland et al. 2005). This
recent version includes about 1000 reactions for $\sim$ 70
molecules containing H, He, C, O, N, Si and S with reaction rates
taken from the UMIST database (Le Teuff et~al. 2000). The general
chemical formalism included in Cloudy has been described in
earlier works \citep[e.g.][]{Hol79, Hol89, Ti85a,Ti85b,LCS91}.
Under low temperature conditions, atomic H may be converted into
H$_2$ (Hollenbach \& McKee 1979). The associative detachment of
H$^-$, may efficiently produce H$_2$ in the high velocity ejecta
via the reaction: H$^-$ + H $\rightarrow$ H$_2$ +e$^-$. Hollenbach
\& McKee (1979, 1989) also suggested that CH formation is possible
via reactions: (i) C + H$_2$ $\rightarrow$ CH + H; (ii) or C$^{+}$
+ H$_2$ + e~$\rightarrow$~CH + H. Both processes could be
effective in this dense component. The first route (i) only needs
the high temperature to activate the reaction. The second route
(ii) is a series of 2 or 3 reactions either at high temperature
(form CH$^{+}$, then form CH$^{2+}$, and then recombine) or has a
small temperature dependence for radiative combination which forms
CH$^{2+}$ directly.

\indent To model the physical conditions in the $-$513 km s$^{-1}$
component we must specify the luminosity, the energy distribution
in ionizing flux, the distance from central source to the ejecta
and elemental abundances in the ejecta.

\indent We have used a model atmosphere flux distribution
\citep{Kur93} corresponding to T$_\mathrm{eff}$=15\,000 K as in
our previous work \citep{VBG05}. We have adopted the distance
$\log$~(R/cm)=17.4 from the central stellar source to the $-$513 km
s$^{-1}$ absorption component (Gull et al. 2005). Independent
studies of elemental abundances in stellar spectra (Hillier et al.
2001), Weigelt Blobs \citep{Duf97,VBG05} and in the S-condensation
beyond the Homunculus (Davidson et al. 1986) demonstrated that He
and N are enhanced by factor 5 and 10 compared to solar values
respectively. While we have adopted He and N abundances, the C and
O abundances are not fixed in the models. Previous attempts to
explain physical conditions by using single density model in a
complex $\eta$~Car environment led to the range for  electron
densities of n$_e$ from 10$^5$ to 10$^{10}$cm$^{-3}$ (Davidson et
al. 1997; Hamann et al. 1999). This large uncertainty has resulted
in the varied hydrogen density in the model from
$\log$\,(n$_\mathrm{H}$/cm$^{-3}$)=5.0 to 9.0.

\indent We have no constraints on how much dust is present in this
ejecta. However, we expect that dust plays an important role in
molecule formation in such a cold environment. Moreover, since
carbon is highly depleted in $\eta$ Car due to CNO-processing, we
expect the grains are predominantly silicates. We have included
Orion Nebula-like silicates in the calculations and varied the
hydrogen density and amount of dust in the model to obtain the
best fit with the observed column densities of
N$_\mathrm{C\,I(tot)}$=2$\times$10$^{14}$ cm$^{-2}$ (Section 3.2),
and N$_\mathrm{Ti \,II(tot)}$=2$\times$10$^{14}$ cm$^{-2}$ (Gull
et al. 2005).

\indent Due to uncertainties in the modeling, including properties
and amount of dust, elemental abundances, thickness of the ejecta
and hydrogen density, we can only constrain an approximate range
of parameters where CH and other molecules form. We have found
that for the density range of
n$_\mathrm{H}\sim$10$^{6-7.5}$cm$^{-3}$, the required total amount
of silicates is 2-10 times larger than that in the Orion Nebula.
At lower density, more dust is needed in the calculations to
reproduce the observed CH column density. Another important
constraint is the H$_2$ column density where we only have a lower
limit as discussed previously.

\indent To demonstrate the sensitivity of molecule formation to
various parameters we present five different, representative
models and information from observations in Table~\ref{models}.
Each model, Table~\ref{models} has the adopted values of C/H and
O/H abundance relative to solar, and includes or not includes dust
and X-ray flux. Table~\ref{models} also gives the calculated shell
thickness ($\Delta$R) given as log($\Delta$R/cm), as well as
average temperature (T$_{ave}$), and predicted logarithmic column
densities of Ti II, C I, $\mathrm{H_2}$, CH, CO, OH and other
important molecules. All models adopt a constant density
n$_\mathrm{H}=$10$^{7}$~cm$^{-3}$, the distance,
$\log(R/cm)$=17.4, from central source to the ejecta, a He
abundance 5 times solar, and N abundance 10 times  solar.
Table~\ref{models} also includes the  deduced values derived from
the species, Ti II, C I, $\mathrm{H_2}$, CH, CO, OH and excitation
temperatures derived from Ti II and CH. A portion of the
ion-molecule chemistry that occurs in the cold molecular cloud may
require a source providing ionization beyond 13.6 eV. In addition
to UV radiation from the primary B star, $\eta$ Car is known to be
a variable in X-rays with a 5.54 year period. The source of these
X-rays is likely due to a binary wind-wind interaction (Corcoran
et al. 2001). We have included an X-ray continuum in the
0.054$-$1keV band of 10$^{37.9}$~erg s$^{-1}$ which is consistent
with the recent identification of He II 4686 (Steiner \& Damineli
2004) and also consistent with the unabsorbed soft X-ray flux seen
in a Chandra spectrum of Eta Car taken on May 3, 2003 (Corcoran et
al. 2005).

\indent In our representative models, the amount of silicates is
fixed at 2.5 times that in the Orion Nebula [dust to gas ratio (by
mass): $\sim$0.004]. In part due to both the presence of grains
and X-ray flux, the local grain-gas photoelectric heating rate
reached $>80\%$ of the total. Models 1 and 2 provide the best fit
to observations (Table~\ref{models}). The C abundance is 0.1 of
solar and O abundance is 0.01 of solar.
 They both contain grains physics (van Hoof et al. 2004), but
Model 2 lacks an X-ray contribution.
 Comparison between Model 1 and Model 2 shows the effects of
 X-ray heating by the increased average temperature, T$_{ave}$.
Yet, there are no significant differences
 in observable column densities.
The only possible discriminant might be the NH column density.

\indent To illustrate effects of different physical processes,
Models 2 and 3 are calculated. Model 3 has solar C and O elemental
abundances with no dust, but includes X-ray flux. The comparison
of Model 3 and Model 1 demonstrates the key role of dust for
H$_2$, CH and CO formation. The predicted molecular column
densities without dust are substantially below that observed. The
presence of dust provides both a site for molecular formation and
a source of pronounced extinction at wavelengths in the far-UV.

\indent Model 4 in Table~\ref{models} demonstrates the effects of
C and O depletion. This model is essentially the same as Model 1,
except for different C and O abundances. In this model, we have
adopted C and O abundances derived from analysis of the Weigelt
Blobs B and D (Verner, Bruhweiler \& Gull 2005), where both C and
O indicate abundances of 0.01$-$0.02 solar. In this model, the
amount of O and C are set at 0.01 of solar. The only observed
molecule containing O is OH. Its predicted column density exhibits
no apparent dependence on C abundance. The only observed molecule
containing C is CH. As expected, the differences between Model 1
and 4 indicate that the CH column density scales with the C
abundance. While the O abundance is found to be the same for the
Weigelt Blobs and for the ejecta, C appears to be enhanced 5-10
times relative to the Blobs (but 0.1 solar) in the $-$513 km
s$^{-1}$ component. If these results reflect a true abundance
differences, one possible explanation could be inhomogeneous
mixing of CNO-processed material with surrounding less synthesized
material in the nebula surrounding $\eta$~Car.

\indent Although Models 1 and 2 provide a basic overall agreement
with the observed column densities or constraints, there are still
several unresolved problems. The elemental gas abundances in these
calculations, except for CNO, are assumed to be solar.
Specifically, no allowances for depletion onto dust grains are
included. If allowances were made, the Ti/C abundance ratio would
be lower, assuming the abundances would follow the trends expected
from condensation temperatures (cf. Lodders 2003). If we were to
adopt a depletion of 0.6 dex, which is that of Si in the diffuse
ISM (Gnacinski 2003), an element with a condensation temperature
close to that for Ti, the agreement with observations would be
good. Still, the predictions for C I, though acceptable, are not
good. Since C I has an ionization potential of 11.26 eV, the
amount of C I is extremely sensitive to both the amount of dust
and the shape of the extinction curve in the far-UV. This
sensitivity makes the C I predictions somewhat uncertain. This
uncertainty also affects the molecular species containing carbon.
The huge difference in observationally derived temperatures of CH
lines (T $\leq$ 30 K) and Ti II lines (T = 760 K) suggests that
there must be temperature and/or density variations within the
filament comprising the $-$513~km~s$^{-1}$ component. Large
temperature differentials between metal and molecules are known in
photoDissociation regions (PDRs; Storzer and Hollenbach 1999).
Absorptions of CH and Ti II arising in such different temperature
regimes cannot be explained by a single density photoionization
model. We speculate that dusty clumps within the filament may
provide an inhomogeneous temperature structure within the
filament. These dusty clumps would provide an efficient way to
spatially differentiate the Ti II and CH regions.

\indent In a CNO-processed gas with C and O subsolar, molecules
with N should be noticeably enhanced. Table~5 shows predicted NH
and NH$_{3}$ column densities, which are smaller than that of CH,
OH and CO. Observations of these molecules would be quite
important in verifying the N-rich
 molecular formation scenario.   Two possibilities include
observations of NH in absorption and NH$_{3}$ in emission. No
spectral lines of NH are located in the region covered by the {\it
HST}/STIS MAMA (1140$-$3160 \AA{}). The strongest NH lines
accessible at optical and UV wavelengths is the R$_{1}$(0) line of
the A$^{3}\Pi-X^{3}\Sigma^{-}$(0-0) band at 3358.053 \AA. However,
this line is quite weak and normally requires very high resolution
and signal-to-noise to detect. For example, in $\zeta$ Oph it has
an equivalent width of $\sim$ 0.4 m\AA~ (Crawford \& Williams
1997). Yet, an enhanced N abundance could improve the chances of
detection. Observations of emission from the rotational
transitions of NH$_{3}$ in the $23.7-24.1$ GHz region offers a
much better possibility (cf. Rizzo et al. 2001; Caproni et al.
2000).

\indent In addition to predicting molecular lines in the ejecta,
the photoionization modeling suggests the presence of the
following strong lines in
 IR: [C I] 1.069 $\mu$m, [O~I] 63.17 $\mu$m, [Si II] 34.81 $\mu$m,
[P II] 136.57 $\mu$m and [Cl~I] 14.34 $\mu$m. These lines should
be investigated in in future high spatial resolution observations for improved understanding
of the excitation conditions, chemical abundances, and dust
physics.

\section{Conclusions} The  {\it HST}/STIS spectra of the
absorption lines produced in the CNO-processed
stellar ejecta around $\eta$ Car show the definite presence of
molecular species in the $-$513~km~s$^{-1}$ velocity component.
The main conclusions are as follows:

1.) The molecules, H$_2$ and CH are positively identified, while
OH is most likely present in the absorption spectrum. All of these
features have velocities within $\pm$ 2~km~s$^{-1}$ of the
previously measured Ti II  $-$513~km~s$^{-1}$ component. The
column densities of observed species are given in the
Table~\ref{models}.\\
2.) The non-detection of CO in the UV is consistent with the
inferred low C and O abundances as found in previous studies. The
referee points out that "any PDR code, even with the C and O
abundances adopted, would give both CH and OH columns greater than
CO.  OH tends to react quickly to form CO, in cases where C is
greater than O in gas phase abundance" and also that "it is easy
to get more CH than CO if the C abundance is greater than O,
but the hard part is to then get more OH than CO."\\
3.)  Using our previous density constraint, 6.0 $\leq$ log
(n$_{H}$/cm$^{-3}$) $\leq$ 7.5, for the $-$513 km~s$^{-1}$
component (Gull et al. 2005) we can produce CH column densities
that are comparable to that in observations. However based upon
our qualitative modeling, we can not rule out the possibility that
the CH is formed at higher densities.\\
 4.) The inadequacies in the modeling can partially be explained by the differences in implied
temperatures as seen in the Ti II and CH with excitation
temperatures of 760 K and less than 30 K, respectively. The
$-$513~km~s$^{-1}$ component may well be inhomogeneous along the
line-of-sight.\\
5.) The modeling is not in complete agreement with observations.
These differences might be explained by temporal variability,
inhomogenieties in temperature and density, or still incomplete
description for chemistry and dust. These problems can only be
resolved by improved modeling and observations revealing other
molecular species providing additional constraints.

\indent The presence of molecules in the ejecta of $\eta$ Car and
our modeling indicates that other molecular species, such as
CH$^+$, CH$_{2}^+$, CH$_{2}$, NH$_{3}$ and NH, may be detectable,
both via absorption and emission at other wavelengths. The
unusually high densities of the gas and the apparent N-rich
environment of the ejecta offer a unique opportunity to probe the
nitrogen-based chemistry in the circumstellar regions of a highly
evolved massive star.

\acknowledgments{ We thank the anonymous referee for useful
comments that improved the paper. The research of EV has been
supported, through NSF grant (NSF - 0206150) to CUA. The
investigators, TRG, EV, KEN, and GV, have been supported through
the {\it HST}/STIS GTO and GO programs. This paper is based upon
observations made with the NASA/ESA Hubble Space telescope,
obtained at the Space Telescope Science Institute, which is
operated by the Association of Universities for Research in
Astronomy, Inc., under NASA contract NAS 5-26555.} HST GO program
9973 was part of the $\eta$ Car Treasury Program, K. Davidson
Principal Investigator.


\clearpage

\begin{deluxetable}{cccccc}
\tabletypesize{\scriptsize} \tablecaption{Data used for the
$\eta$\,Car $-$513 km\,s$^{-1}$ Ejecta Analysis. The central
wavelengths for the E140M and the E230H gratings are 1425 and 3012
\AA{}, respectively. \label{obs}} \tablewidth{0pt} \tablehead{
\colhead{HST GO}  & \colhead{Observing}  & \colhead{JD}  &
\colhead{Grating} & \colhead{Spectral} &
\colhead{Analyzed} \\
\colhead{Program} & \colhead{Date} & \colhead{(245\,0000+)} &
\colhead{} & \colhead{Coverage (\AA)} & \colhead{ Species}
 }
\startdata
9083 & 2002$-$01$-$20 &  2294.735 & E140M & 1150-1729 & H$_2$,  C
I, CO \\
9083 & 2002$-$01$-$20 &  2294.831 & E230H & 2886-3158 & OH, CH \\
9973 & 2003$-$09$-$21 &  2904.012 & E230H & 2378-3158 & OH, CH \\
\enddata
\end{deluxetable}

\clearpage

\begin{deluxetable}{lcccccc}
\tabletypesize{\scriptsize} \tablecaption{Identified CH Lines in
the $\eta$\,Car $-$513 km\,s$^{-1}$ ejecta. All transitions are
from the C$^{2} \Sigma^{+}$ $-$ X$^{2}\Pi$ system, 0$-$0 band
($\sigma_0$=31778 cm$^{-1}$). \label{CH}} \tablewidth{0pt}
\tablehead{ \colhead{$\lambda_\mathrm{lab}$}  &
\colhead{J$_\mathrm{l}$} & \colhead{E$_\mathrm{l}$} &
\colhead{log\,$gf$\tablenotemark{1}} &
\colhead{$\lambda_\mathrm{obs}$}  & \colhead{Vel.}  &
W$_\lambda$\\
\colhead{(\AA)} & \colhead{ } & \colhead{(cm$^{-1}$)} & \colhead{
} & \colhead{(\AA)} & \colhead{(km\,s$^{-1}$)} &
\colhead{(m\AA{})} } \startdata
3144.079  &  1/2  &     0.00  & $-$2.28& 3138.678 & $-$515.0 &
6.3$\pm$2.0\tablenotemark{2}\\
3144.102  & 1/2  &     0.00  & $-$1.98& 3138.722 &$-$513.0 & \\
3146.922  &  1/2  &     0.04  & $-$1.98& 3141.525 &$-$514.1 &\\
3145.836  & 3/2  &     17.77 & $-$1.53&3140.459 & $-$512.6
&6.0$\pm$2.0\\
3140.221  &  3/2  &     17.81 & $-$1.93&3134.852 & $-$512.6 & \\
3148.683  & 3/2  &     17.81 & $-$1.70&3143.302 & $-$512.3 &
7.0$\pm$2.0\\
3145.071  &3/2  &   66.84 & $-$1.53& &  &  $<$ 2 \\
3150.749  &  3/2  &   67.11 & $-$1.71& &  &$<$ 3 \\
3145.679  &5/2  &   73.07 & $-$1.34&  & &  $<$ 2 \\
3151.330  &5/2  &  73.19  & $-$1.55&  & &  $<$ 2.4 \\
\enddata
\tablecomments{Values are based on January 2002 data, E230H.}
\tablenotetext{1}{http://kurucz.harvard.edu/LINELISTS/}
\tablenotetext{2}{$\sum$W$_\mathrm{CH}$ for CH
$\lambda\lambda$3144.079, 3144.102}
\end{deluxetable}

\clearpage

\begin{deluxetable}{lcccccc}
\tabletypesize{\scriptsize} \tablecaption{OH lines in the
$\eta$\,Car $-$513 km\,s$^{-1}$ Ejecta from the A   $^{2}
\Sigma^{+}$ $-$ X$^{2}\Pi_{3/2}$ System, 0$-$0 band \label{OH}}
\tablewidth{0pt} \tablehead{ \colhead{$\lambda_\mathrm{lab}$}  &
\colhead{J$_\mathrm{l}$} & \colhead{E$_\mathrm{l}$} &
\colhead{log\,$gf$\tablenotemark{1}} &
\colhead{$\lambda_\mathrm{obs}$}  & \colhead{Vel.}  &
W$_\lambda$\\
\colhead{(\AA)} & \colhead{ } & \colhead{(cm$^{-1}$)} & \colhead{
} & \colhead{(\AA)} & \colhead{(km\,s$^{-1}$)} &
\colhead{(m\AA{})} } \startdata
3082.559  &  3/2  &     0.00  & $-$2.59& 3077.262 & $-$515.1 &
2.3$\pm$1.0\\
3079.36\tablenotemark{2}  & 3/2  &     0.00  & $-$2.38& 3074.093
&$-$512.8 & 1.9$\pm$1.0\\
\enddata
\tablecomments{Values are based on January 2002 data, E230H.}
\tablenotetext{1}{Roueff
E. (1996) MNRAS 279, L37} \tablenotetext{2}{$\sum$W$_\mathrm{CH}$
for OH
$\lambda\lambda$3079.36,3079.23; gf-value reflect sum of both
lines.}
\end{deluxetable}

\clearpage

\begin{deluxetable}{lcccccc}
\tabletypesize{\scriptsize} \tablecaption{Measured  \ion{C}{1}
Lines in the $\eta$\,Car $-$513 km\,s$^{-1}$ Ejecta.\label{CI}}
\tablewidth{0pt} \tablehead{ \colhead{$\lambda_\mathrm{lab}$}  &
\colhead{J$_\mathrm{l}$} & \colhead{E$_\mathrm{l}$} &
\colhead{log\,$gf$\tablenotemark{1}} & \colhead{W$_\lambda$}  &
\colhead{N$_\mathrm{C\,I}$}\\
\colhead{(\AA)} & \colhead{ } & \colhead{(cm$^{-1}$)} & \colhead{
} & \colhead{(m\AA)} & \colhead{(cm$^{-2}$)} } \startdata
1270.143  & 0          & 0.00  & $-$3.41 & $<$3 & \\
1280.135  & 0          & 0.00 & $-$1.58 & 18.6: & \\
1328.833  & 0          & 0.00 & $-$1.34 & 29.8$\pm$4& 13.8\\
1560.309  & 0          & 0.00 & $-$1.11 & 58.8:& \\
1656.928  & 0          & 0.00 & $-$0.83 & 54.4:& \\
1329.578  & 1          & 16.40 & $-$0.64 & 49.5:&\\
1656.267  & 1          & 16.40 & $-$0.73 & 48.3:& \\
1657.379  & 1          & 16.40 & $-$0.95 & 23.5$\pm$4& 13.6\\
1658.128  & 2          & 43.40 &  $-$0.73 &  38.7$\pm$4& 14.0\\
1661.438  & 2          & 43.40 & $-$0.49 & 60.1:& \\
\enddata
\tablenotetext{1}{Morton, D.C. (2003), ApJS, 149. 205}
\end{deluxetable}

\clearpage

\begin{deluxetable}{cccccc}
\tabletypesize{\scriptsize}
\tablecaption{Comparison between Observations and Models for the
$\eta$\,Car $-$513 km\,s$^{-1}$ Ejecta. \label{models}}
\tablewidth{0pt} \tablehead{ \colhead{ } & \colhead{Observations}
& \colhead{ } & \colhead{ } & \colhead{Models\tablenotemark{1}} &
\\
\colhead{ } & \colhead{ } & \colhead{1} & \colhead{2} &
\colhead{3} & \colhead{4}}

\startdata
  T$_\mathrm{av}$(K) &  & 993 & 572 & 1120 & 1010 \\
  T$_\mathrm{TiII}$(K) & 760 &  &  &  \\
  T$_\mathrm{CH}$(K) & 30 &  &  &  \\
  Log($\Delta$ R/cm) &  & 14.8 & 14.8 & 14.4 & 14.8 \\
  C &  &   0.1   &  0.1    &  1    &    0.01  \\
  O &  &   0.01   &  0.01  &  1    &  0.01   \\
  X-ray & & yes &  no & yes & yes \\
  Dust & & yes & yes & no & yes \\
  Log(N$_\mathrm{H_2}$/cm$^{-2}$) & $>$~16 & 20.7 & 20.7 &
13.7 & 20.7 \\
  Log(N$_\mathrm{C I}$/cm$^{-2}$) & 14.3 & 14.9 & 14.9 &
14.1 & 13.9 \\
  Log(N$_\mathrm{Ti II}$/cm$^{-2}$) & 14.3 & 14.8 & 14.8 &
14.5 & 14.8 \\
  Log(N$_\mathrm{CH}$/cm$^{-2}$) & 13.7 & 13.2 & 13.2 & 6.7
& 12.3 \\
  Log(N$_\mathrm{OH}$/cm$^{-2}$) & $13.5$  & 13.8 & 13.7 &
10.4 & 13.8 \\
  Log(N$_\mathrm{CO}$/cm$^{-2}$) & $\leq$ ~12.95 & 12.7 & 12.6 & 8.3 & 11.8 \\
  Log(N$_\mathrm{CH^+}$/cm$^{-2}$) &  & 12.2 & 12.3 & 10.1 &
11.3 \\

  Log(N$_\mathrm{CH_2}$/cm$^{-2}$) &  & 12.5 & 12.6 &  & 11.6
\\
  Log(N$_\mathrm{NH}$/cm$^{-2}$) &  & 11.9 & 11.3 & 9.0 &
11.8 \\
  Log(N$_\mathrm{NH_3}$/cm$^{-2}$) &  & 8.2 & 5.6 &  &
8.2 \\
\\\enddata
\tablenotetext{1}{All models are calculated at $\log$(R/cm)=17.4
and $\log$(n$_\mathrm{H}$/cm$^{-3}$)=7. The thickness of the
cloud, $\Delta R$, is in cm, the column densities are listed. The
C and O abundances are presented relative to solar values. X-ray
flux is based on $\it{Chandra}$ observations (Corcoran et al.
2005). Dust assumes the Orion silicates dust with the amount 2.5
times larger than that in the Orion Nebula.  We have indicated
whether the dust or and X-rays are included in the model by "yes"
and "no". See text for more details. Predictions on other
molecules are available upon request at kverner@fe2.gsfc.nasa.gov}
\end{deluxetable}

\clearpage

\figcaption [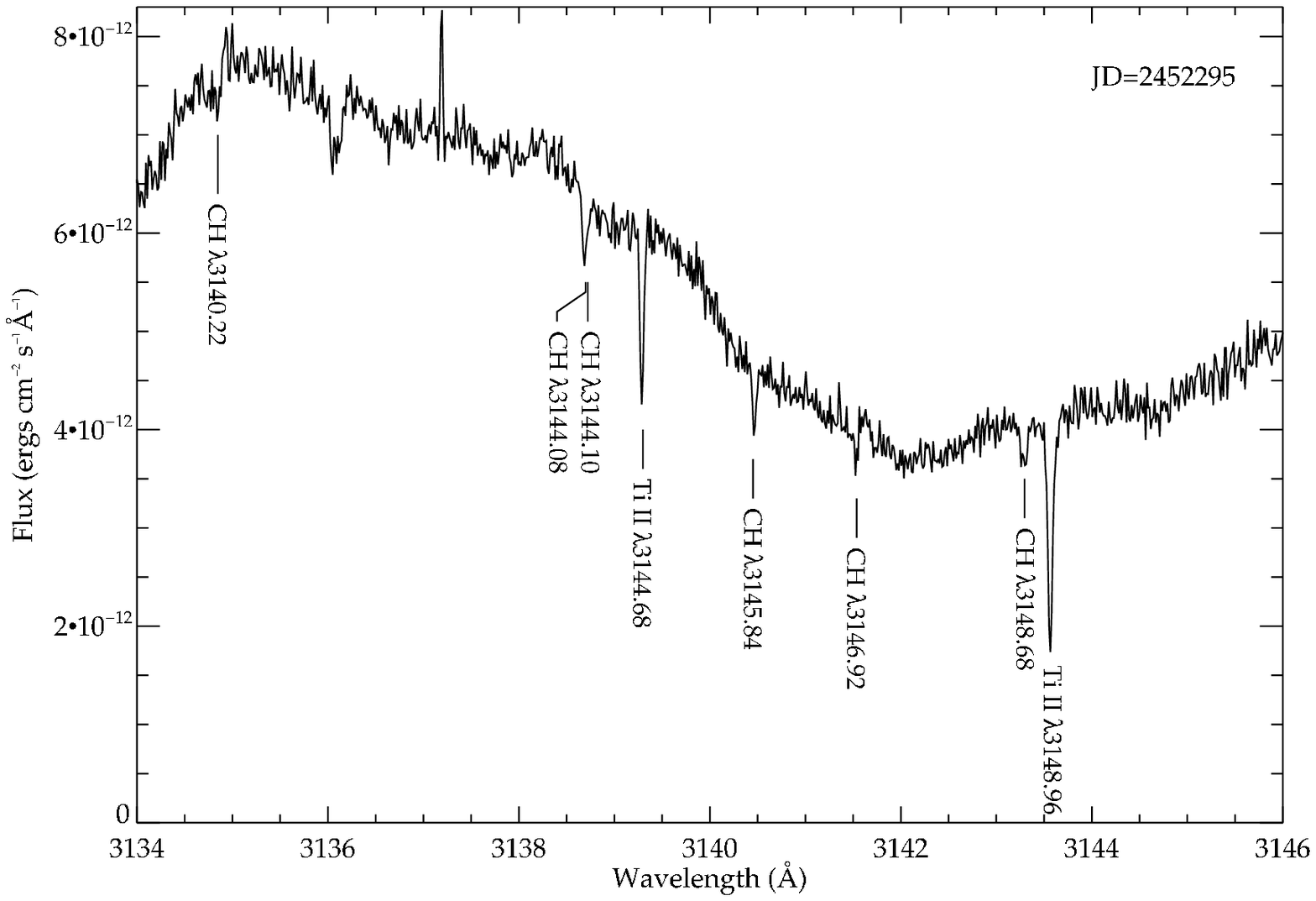]{The spectrum in 3137$-$3145 \AA{} range
obtained with the {\it HST}/STIS  in January 2002 (JD=245\,2294).
The positions of six CH and two strong Ti II lines are marked by
solid lines and their laboratory wavelengths.}

\figcaption [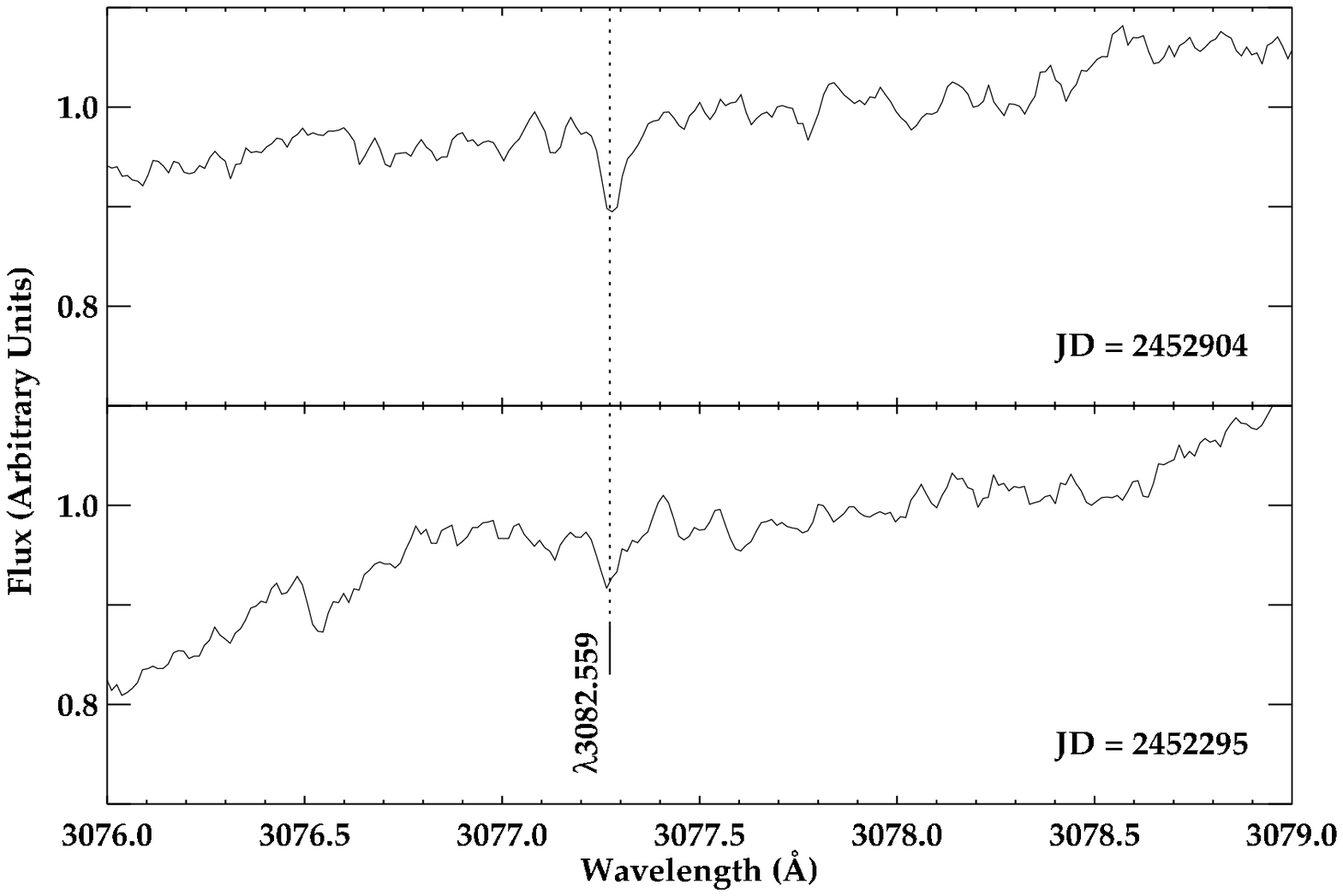]{The spectra in 3076$-$3079 \AA{} range
obtained with the {\it HST}/STIS  in January 2002 and in September
2003. The position of the OH line is marked by solid line and its
laboratory wavelength in each spectrum.}

\figcaption [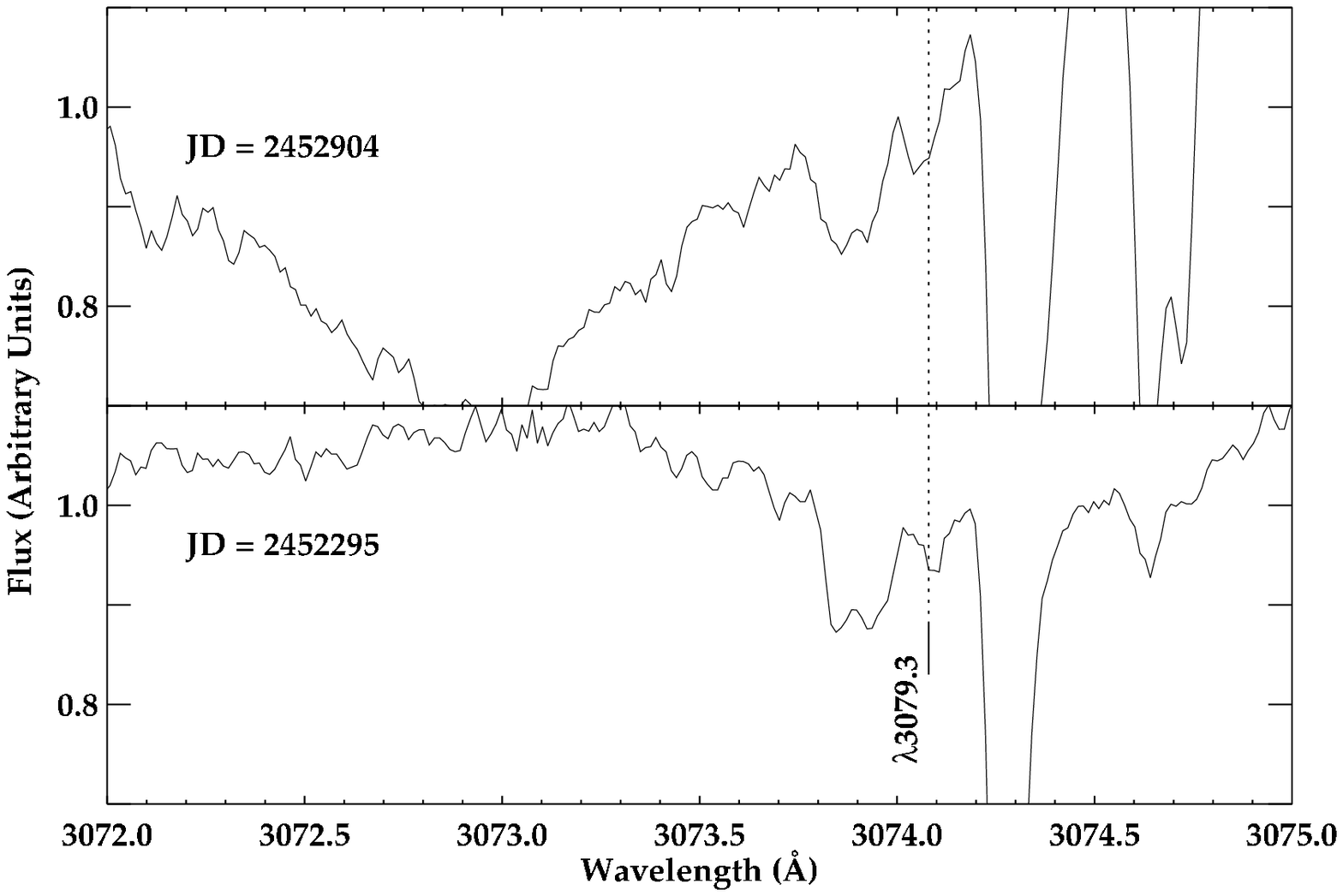]{The spectra in 3072$-$3075 \AA{} range
obtained with the {\it HST}/STIS  in January 2002 and in September
2003. The position of the OH line is marked by solid line and its
laboratory wavelength in each spectrum.}

\figcaption [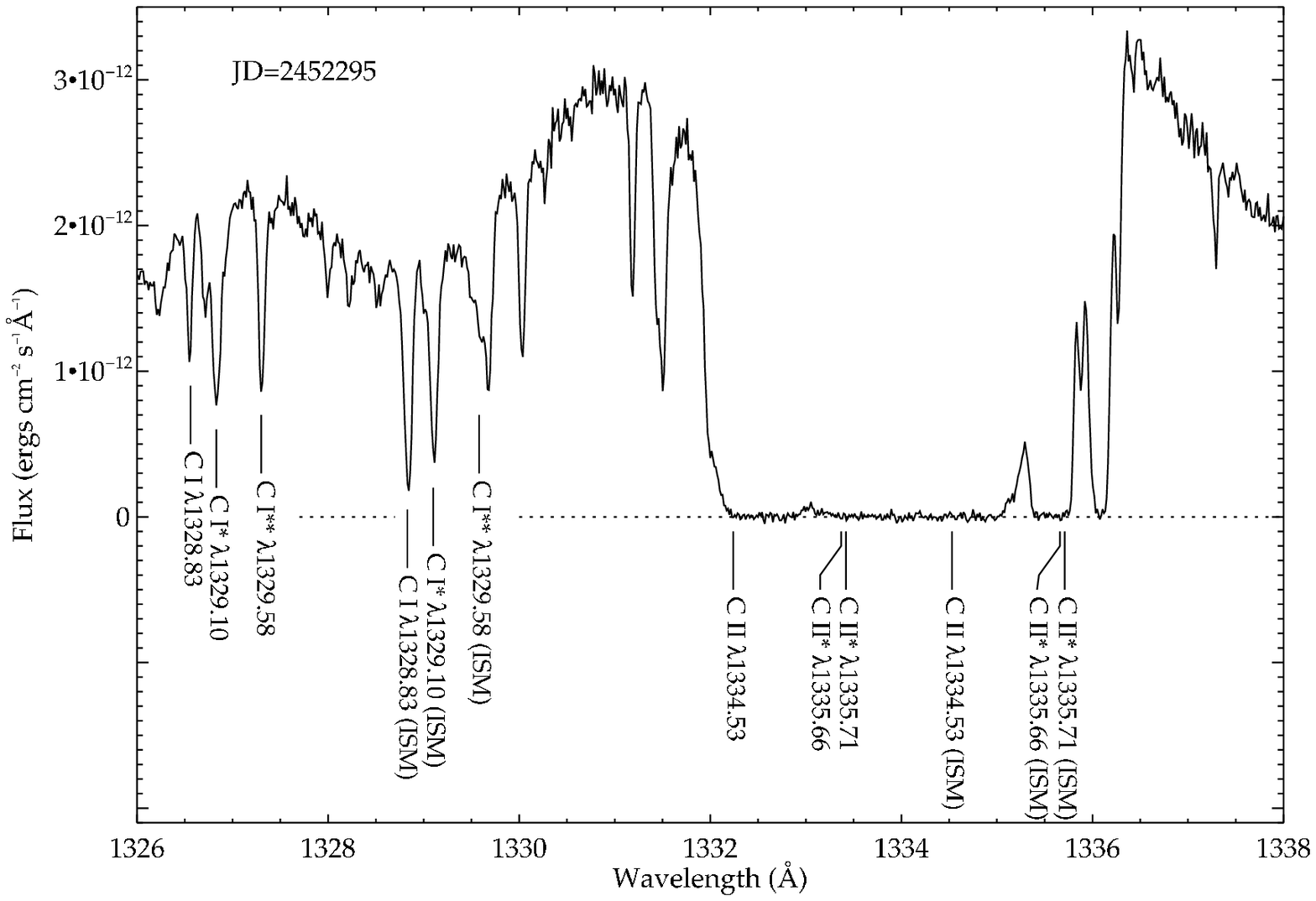]{The spectrum in 1326$-$1338 \AA\ range
obtained with {\it HST}/STIS  in January 2002 (JD = 2452295). The
position of the detected C I and possible C II lines are marked by
solid lines and laboratory wavelengths.  The expected presence of
C II in the fast ejecta is completely obscured by the broad,
saturated  C II $\lambda$1335 stellar wind profile in the same
region. Positions of C I and C II lines originated in interstellar
medium (ISM) are plotted for comparison.}

\newpage
\rotatebox{00}{\epsscale{1.0} \plotone{f1.ps}}
\newpage
\rotatebox{00}{\epsscale{1.0} \plotone{f2.ps}}
\newpage
\rotatebox{00}{\epsscale{1.0} \plotone{f3.ps}}
\newpage
\rotatebox{00}{\epsscale{1.0} \plotone{f4.ps}}
\end{document}